\documentclass[12pt,english,floatfix,preprint,amsmath,amssymb,aps,prd,superscriptaddress, titlepage]{revtex4-2}

\usepackage{graphicx}
\usepackage[utf8]{inputenc}
\usepackage[T1]{fontenc} 
\usepackage{microtype} 
\usepackage{physics}    
\usepackage{tensor}
\usepackage{tensor} 
\usepackage[svgnames]{xcolor} 
\usepackage[obeyFinal, 
    color       = LightGray,
    bordercolor = LightGray,
    textsize    = footnotesize,
    figwidth    = 0.99\linewidth,
    prependcaption]{todonotes} 
\usepackage[allcolors=blue,colorlinks,pdftitle={Shadow and Quasinormal Modes of a Radiating Compact Object},pdfauthor={Caio C. Rodrigues and A. C. L. Santos}]{hyperref} 
\usepackage{orcidlink} 

\begin{document}
\title{Echoes and quasinormal modes of asymmetric black bounces}
\author{A. C. L. Santos\,{\orcidlink{0000-0002-0080-5699}}}
\email{alanasantos@fisica.ufc.br}
\affiliation{Universidade Federal do Ceará (UFC), Departamento de Física,
 Campus do Pici, Fortaleza- CE, C.P. 6030, 60455-760- Brazil}
\affiliation{Instituto de Física Corpuscular (IFIC), CSIC‐Universitat de Val\`{e}ncia, Spain} 
\author{L. A. Lessa\, {\orcidlink{0009-0009-1961-9819} }}
\email{leandrophys@gmail.com}
\affiliation{Programa de Pós-graduação em Física, Universidade Federal do Maranhão, Campus Universitário do Bacanga, São Luís (MA), 65080-805, Brazil}
\author{R. V. Maluf\,{\orcidlink{0000-0002-9952-4589}}}
\email{r.v.maluf@fisica.ufc.br}
\affiliation{Universidade Federal do Ceará (UFC), Departamento de Física,
 Campus do Pici, Fortaleza- CE, C.P. 6030, 60455-760- Brazil}
\author{G. J. Olmo\,{\orcidlink{0000-0001-9857-0412}}}
\email{gonzalo.olmo@uv.es}
\affiliation{Instituto de Física Corpuscular (IFIC), CSIC‐Universitat de Val\`{e}ncia, Spain}  
\affiliation{Universidade Federal do Ceará (UFC), Departamento de Física,
 Campus do Pici, Fortaleza- CE, C.P. 6030, 60455-760- Brazil}

\date{\today}

\begin{abstract}
We study quasinormal modes and echoes of symmetric and asymmetric black bounce solutions generated by anisotropic fluids within the framework of general relativity. We derive the effective potential governing massless scalar fields and compute the corresponding quasinormal mode spectra using three independent methods: sixth-order WKB, Pöschl–Teller and time-domain evolution. Our results show that symmetric black bounce configurations with horizons yield a standard single-barrier potential, while horizonless solutions {may} exhibit multiple potential barriers that generate gravitational wave echoes. These echoes are sensitive to model parameters such as the fluid energy density and the regularizing parameter $a$ that defines the minimal $2-$sphere. The asymmetric models considered recover the Reissner–Nordström solution in their external region but can be bounded or unbounded in the inside, depending on the sign of a parameter. Both cases have similar qualitative properties as far as wave emission is concerned but show no echoes. This makes it very difficult to distinguish them from standard Reissner–Nordström configurations. 
\end{abstract}

\maketitle

\section{Introduction}\label{sec: intro}
One of the main open questions in theoretical physics is understanding the behavior of gravity in strong-field scenarios. Despite the experimental success of general relativity (GR) \cite{Will:2014kxa}, the conceptual challenges posed by singularities \cite{Penrose:1964wq,Geroch:1968ut} in scenarios involving natural matter sources suggest that the theory should be improved in some way, leaving observable imprints at some scales. This {\it million dollar question} has stimulated a good deal of activity in the theoretical physics community in many diverse fronts. In this sense, recent years have seen a rise in phenomenological models aiming to ``resolve'' these singularities within a classical framework, without relying on prior knowledge of the underlying mechanisms of a potential quantum theory of gravity~\cite{Malafarina:2017csn,Kelly:2020lec,Bambi:2013caa}. Among the most widely investigated models in this regard,  \textit{black bounces} \cite{Simpson:2018tsi} appear as a reasonable alternative to traditional black holes, and even may emerge in non-perturbative quantum gravity approaches \cite{Muniz:2024wiv}. 

The term {\it black bounce} was first coined in a seminal paper by Visser and Simpson~\cite{Simpson:2018tsi}. These objects are characterized by smoothly interpolating between nonsingular black holes and wormholes, depending on a single free parameter that can be tuned at will. Specifically, for spherically symmetric configurations, the area of the 2-spheres is constrained to have a lower bound or ``bounce''. When this minimal sphere is hidden behind an event horizon, one speaks of a {\it black bounce}~\cite{Lobo:2020ffi,Mazza:2021rgq}. These objects are (usually) free from singularities because the bound in the radial sector keeps curvature invariants finite and, crucially, allows geodesic congruences to reexpand beyond the throat, guaranteeing geodesic completeness~\cite{LimaJunior:2025uyj,Carballo-Rubio:2019fnb}.


Since the areal function is key to keeping curvature invariants in these models within finite bounds, it is important to understand how the matter sources contribute to shape the radial sector of the metric. In \cite{Bronnikov:2021uta} it was shown that the Simpson-Visser metric is supported by a matter content consisting of nonlinear electrodynamics plus a phantom scalar field, with the electromagnetic source corresponding to a magnetic charge. Interestingly, these solutions can also be replicated by considering an electrically charged source \cite{Alencar:2024yvh}. Furthermore, the relation between these models and modified gravitational dynamics has been explored in \cite{Alencar:2024nxi, Junior:2024vrv, Junior:2024cbb, Atazadeh:2023wdw, Fabris:2023opv}. Their impact on various physical observables, including gravitational lensing \cite{Nascimento:2020ime, Cheng:2021hoc, Islam:2021ful, Ghosh:2022mka}, black hole shadows \cite{Guerrero:2021ues, Olmo:2023lil, Guerrero:2022qkh}, and gravitational wave echoes \cite{Silva:2024fpn, Ou:2021efv, Yang:2021cvh}, among other phenomena \cite{Franzin:2022iai, Zhang:2022zox, Yang:2022ryf, Javed:2023iih, daSilva:2023jxa}, have also been extensively investigated. In all these works, the  condition $g_{tt}=g_{rr}^{-1}$ was assumed. However, new classes of black bounce solutions can be derived by relaxing this condition, as demonstrated recently in Ref. \cite{Lessa:2024erf}. These solutions are supported by an anisotropic fluid that adheres to certain simplifying constraints on the matter content. Specific examples of such fluids are nonlinear theories of electrodynamics. 


Leaving aside theoretical considerations and focusing on observational aspects, the recent detection of gravitational waves by LIGO/Virgo from the mergers of binary black holes~\cite{LIGOScientific:2016aoc,LIGOScientific:2016sjg,LIGOScientific:2017bnn,LIGOScientific:2017ycc} and binary neutron stars~\cite{LIGOScientific:2017vwq} has significantly heightened the relevance of studying gravitational wave emissions from massive compact objects. The ringdown phase of such mergers can be described using perturbation theory and is characterized by a set of complex frequencies known as quasinormal modes (QNMs), where the real part is associated with the oscillation frequency and the imaginary part with the damping rate. Interestingly, these complex frequencies depend essentially on the parameters that characterize the interaction between the spacetime and the corresponding test field — such as mass, charge and angular momentum \cite{Berti:2009kk}. For this reason, the computation of QNMs has also become a focus of intense research \cite{Konoplya:2025hgp, Volkel:2018hwb,Magalhaes:2023xya, Konoplya:2025mvj,Alfaro:2024tdr}, including black bounce scenarios \cite{Yang:2022ryf, Wu:2022eiv, Ou:2021efv, Guo:2021wid}. In particular, investigations of scalar perturbations for a background described by the rotating Simpson-Visser solution can be found in Ref.~\cite{Franzin:2022iai}. Any deviation from the classical black-hole paradigm is expected to leave a distinct imprint on the QNM spectrum and, consequently, provide a means to distinguish between different spacetime configurations, as demonstrated in wormhole-like models featuring multiple potential barriers and gravitational wave echoes \cite{Bambi:2025wjx}. The spectroscopic characterization of compact objects is thus a line of research that is gaining momentum in recent years \cite{Volkel:2025jdx,Berti:2025hly,Furuta:2024jpw,Moreira:2023cxy,Cardoso:2019rvt,Dreyer:2003bv}. 

In this sense, the main goal of this work is to examine the ringdown phase of the previously mentioned exact compact object solutions in the context of general relativity \cite{Lessa:2024erf}, focusing on the propagation of linear test fields, which describe scalar perturbations minimally coupled to the metric. We compute the QNM spectrum using three independent methods —sixth-order WKB (Wentzel–Kramers–Brillouin) expansion, Pöschl–Teller approximation, and time-domain integration— finding good agreement among the results. In addition, we investigate the possible emergence of gravitational wave echoes arising from the structure of the effective potential in horizonless configurations.

In particular, we will consider two different types of black bounces, one representing a standard symmetric object, built in terms of the Kiselev solution and replacing $r\to \sqrt{a^2+r^2}$, and another illustrating an asymmetric bounce, which may transition to a bounded universe or to an unbounded universe, depending on the sign of a parameter. The relevance of this second example, which we will compare with a Reissner-Nordström geometry, lies in the fact that one can have three different geometries with similar properties away from the minimal 2-surface but completely different internal regions. Our goal is to determine if such different internal structures can be clearly identified by the observation of gravitational waves emission during the ringdown phase. As we will see, the interaction of waves with the internal regions not always allow to clearly  distinguish them from other structures, such as a standard Reissner-Nordström configurations (naked or not).  

Our paper is organized as follows: In Sec. \ref{secII}, we introduce the symmetric and asymmetric black bounce geometries derived from GR coupled to an anisotropic fluid. Sec. \ref{secIII} presents the framework for analyzing scalar perturbations in these backgrounds, including the derivation of the effective potential and the description of the semi-analytical and numerical methods employed. In Sec. \ref{secIV}, we discuss our results, highlighting the differences between configurations with and without horizons. Finally, we outline our perspectives and conclusions in Sec. \ref{secV}.

\section{SYMMETRIC AND ASYMMETRIC BOUNCE MODELS}\label{secII}

Since our main interest is the study of the propagation of test scalar fields (perturbations) on the novel black bounce spacetimes constructed in Ref. \cite{Lessa:2024erf}, first we will present a brief summary of such geometries. The metric can be written as
\begin{equation} \label{metricgeral}
ds^2 = -A(r)dt^2 + B(r)dr^2 + \Sigma(r)^2 d\Omega^2,
\end{equation}
and its dynamics is governed by the Einstein field equations, namely,
\begin{equation}\label{eintein}
R_{\mu\nu} - \frac{1}{2} R g_{\mu\nu} = \kappa^2 T_{\mu\nu},
\end{equation}
where $T_{\mu\nu}$ is the stress-energy tensor, and $\kappa^2$ represents Newton's constant, which we set to unity for simplicity. The area function $\Sigma(r)$ is assumed to be non-monotonic and having a minimum, which signals the location of the compact object's throat (or bounce). This minimality condition can be mathematically expressed as
\begin{equation}
    \Sigma(r_0)\neq 0 \ \  , \ \ \\  \Sigma'(r_0) =0 , \ \ \\  \Sigma''(r_0) >0 
\end{equation}
where $r_0$ denotes the radial coordinate of the throat. On the other hand, the presence of event horizons can be encoded in the zeros of the metric function $A(r)$, and they need not have any {\it a  priori} relation with $r_0$. The specific form of the area function $\Sigma(r)$ and the metric functions $A(r)$ and $B(r)$ are determined by the assumed matter content. As demonstrated in Ref.\cite{Lessa:2024erf}, within the framework of GR, it is possible to generate both symmetric and asymmetric black bounce solutions for an anisotropic fluid, i.e., $T^{\mu}{}_{\nu} =\text{diag}(-\rho,p_r,p_t,p_t)$, where $\rho$, $p_r$, and $p_t$ represent the energy density, the radial pressure and the tangential pressure of the fluid, respectively, with the following equations of state: $\rho+p_r=0$ and $p_t = \omega\rho$, where $\omega$ is a dimensionless constant. Note that nonlinear electrodynamics models satisfy these conditions.  Under these assumptions, and considering the conservation of the stress-energy tensor (i.e., $\nabla_{\mu}T^{\mu\nu}=0$), it is straightforward to show that the area function is given by
\begin{equation}\label{sigmageral}
    \Sigma(r)={\Sigma_0}/{\rho(r)^{\frac{1}{2(\omega+1)}}} \ ,
\end{equation}
where $\Sigma_0$ is an integration constant. If the effective energy density of the fluid is bounded from above, the radial function $\Sigma(r)$ will automatically develop a nonzero minimum, which represents a throat or bounce. On the other hand, the determination of the metric functions $A$ and $B$ follows from the Einstein field equations~\eqref{eintein}. For instance, the condition $\rho+p_r=0$ ensures that $B(r) = \frac{\Sigma'(r)^2}{A(r)}$, allowing the line element~\eqref{metricgeral} to be expressed in a more convenient form:
\begin{equation}\label{metricA}
    ds^2 = - A(\Sigma)dt^2 + \frac{d\Sigma^2}{A(\Sigma)}+\Sigma^2 d\Omega^2,
\end{equation}
where we have adopted the area function $\Sigma(r)$ itself as a radial coordinate. The function $A$ must satisfy the equation
\begin{equation}\label{eq:AofSigma}
    \Sigma A_{\Sigma} + A = 1 - \rho \Sigma^2 \ ,
\end{equation}
where we are denoting $A_{\Sigma}= \frac{\partial A}{\partial \Sigma}$. However, as pointed out in Ref. \cite{Lessa:2024erf}, the line element~\eqref{metricgeral} contains more information than~\eqref{metricA} because using $\Sigma$ as a radial coordinate is only valid in domains where $d\Sigma/dr\neq 0$. Though the representation \eqref{metricA} is very compact and convenient, it is unable to see the extrema of $\Sigma(r)$, which contain crucial physical information about the matter and the geometry. Therefore, given a matter density profile $\rho(r)$ that satisfies the previously mentioned equations of state, we obtain the line element~\eqref{metricgeral} with the function $\Sigma(r)={\Sigma_0}/{\rho(r)^{\frac{1}{2(\omega+1)}}}$ providing complete information about the dependence of the 2-spheres on the radial coordinate $r$. In what follows, we will present some concrete black bounce solutions and will subsequently study their quasi-normal modes spectra.


\subsection{Model I}
The symmetric black bounce solution proposed in   \cite{Lessa:2024erf} is supported by an energy density of the form
\begin{equation}\label{energi}
    \rho (r)= \frac{\rho_0}{(a^2+r^2)^{1+\omega}}
\end{equation}
where $\rho_0$ and $a$ are constants.  Substituting this expression into Eq.~\eqref{sigmageral}, we obtain the areal function given by 
\begin{equation}
\Sigma_I(r) =  \sqrt{a^2+r^2},    
\end{equation}
which exhibits the regular minimum $\Sigma_I(0) = a$ at $r=0$ and leads to the line element
\begin{equation}\label{symetric}
ds^2 = - A_I(r)dt^2 + \frac{1}{\left(1 + \frac{a^2}{r^2}\right)A_I(r)}dr^2 + (a^2 + r^2)  d \Omega^2,  
\end{equation}
where,
\begin{equation}
A_I(r) = 1 - \frac{2m}{\sqrt{a^2 + r^2}} + \frac{\rho_0}{(2\omega - 1)(a^2 + r^2)^\omega}.    
\end{equation}
In the limit $a=0$, this solution recovers the Kiselev metric \cite{Kiselev:2002dx}, which describes a black hole surrounded by an anisotropic fluid (see \cite{Visser:2019brz} for obvious clarifications). For 
$\omega=1$, the function $A_I(r)$ coincides with the Reissner-Nordström case under the replacement $r\to \sqrt{a^2 + r^2}$, where $\rho_0$ can be interpreted as an effective charge. The resulting geometry is regular everywhere. Regarding the horizons, one finds that they are located at
\begin{equation}
    r_h = S_1 \sqrt{\left( 1 + S_2 \sqrt{1 - 4m\rho_0} \right)^2 \rho_0^{-2} - a^2},
\end{equation}
with $S_1, S_2 = \pm 1$. Depending on the parameters, we can have a regular black hole, an extreme black hole, a traversable wormhole or a hyperextremal configuration.

\subsection{Model II}

\begin{figure}[!h]
\includegraphics[height=6.0cm]{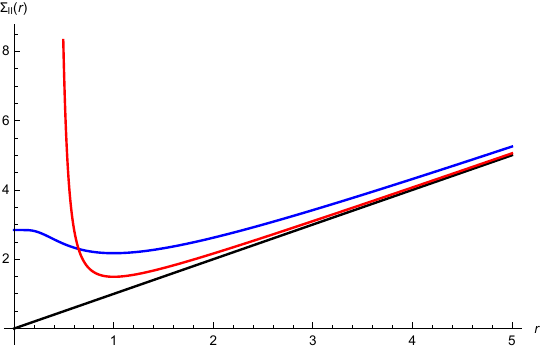} 
\caption{Graphical representation of $\Sigma_{II}(r)$ as a function of $r$ with $r_0 = 1$, $l_0=+1$ (blue) and $l_0=-1$ (red) with the normalization given in (\ref{eq:norm}) for $\alpha=1$. The black line represents $\Sigma_{II}(r)=r$. \label{fig:Sigma2improved}}
\end{figure}

A non-standard asymmetric black bounce geometry was built in \cite{Lessa:2024erf} using the area function 
\begin{equation}
\Sigma_{II} (r) = \tilde{\Sigma}_0 e^{-\alpha^2 \chi(r)},    
\end{equation}
where,
\begin{equation}
\chi(r) \equiv \text{Ei} \left(-\frac{l_0}{r} \right) + \frac{r_0}{l_0} e^{-\frac{l_0}{r}},
\end{equation}
$r$ is defined only in the interval $r > 0$, Ei(z) represents the exponential integral function and $\alpha, \tilde{\Sigma}_{0}$ and $l_{0}$ are constants. This function arises as a solution of the differential equation 
\begin{equation}
\frac{\Sigma_r}{\Sigma}=\frac{\alpha^2 e^{-\frac{l_0}{r}}}{r}\left(1-\frac{r_0}{r}\right) \ ,    
\end{equation}
which clearly possesses and extremum at $r=r_0$ (for any sign of $l_0$). For our purposes, it is convenient to normalize $\tilde{\Sigma}_0$ as
\begin{equation}\label{eq:norm}
\tilde{\Sigma}_0\equiv \left(|l_0|e^{\frac{l_0 \gamma+\text{r0}}{l_0}}\right)^{\alpha^2} \ ,  
\end{equation}
where $\gamma$ denotes the Euler-Mascheroni constant, take $\alpha=1$ to guarantee that $\Sigma(r)$ grows linearly with $r$ at infinity, and also consider a constant shift of the form $\Sigma_{II} (r)\to \Sigma_{II} (r)-(l_0+r_0)$ to guarantee that in the far limit we do recover $\Sigma_{II} (r)=r+O(1/r)$. To be explicit, we will be dealing with the area function
\begin{equation}
\Sigma_{II} (r) = \left(|l_0|e^{\frac{l_0 \gamma+\text{r0}}{l_0}}e^{-\chi(r)}\right)^{\alpha^2}-(l_0+r_0),    
\end{equation}
which has the same extrema as the original one presented in \cite{Lessa:2024erf} but with improved asymptotics. This will facilitate the comparison of different models, guaranteeing that sufficiently large values of $r$ describe the same $2-$spheres (see Fig.\ref{fig:Sigma2improved}). For this purpose, we must always assume that $\alpha=1$.

The asymmetry of these solutions below the minimum (at $r=r_0$) is evident from Fig.\ref{fig:Sigma2improved}, with the blue curve ($l_0>0$) representing a bounded internal region (with the area of the $2-$spheres going to a constant as $r\to 0$), while the red curve ($l_0<0$) representing a wormhole (with the area of the $2-$spheres diverging as $r\to 0$). The energy density of these objects can be obtained by using the relation (\ref{sigmageral}) in the form  
\begin{equation}
\rho(r) = \tilde{\rho}_0 \left( \frac{\tilde{\Sigma}_0}{\Sigma(r)} \right)^{2(\omega+1)} \ ,  
\end{equation}
and, as expected, its maximum lies at $r=r_0$. The complete metric is described by
\begin{equation}\label{asymmetric}
ds^2 = -A_{II}(r)dt^2 + B_{II}(r)dr^2 + \Sigma_{II}(r)^2 d\Omega^2,
\end{equation}
with,
\begin{equation}
A_{II}(r) =
\begin{cases}
1 - \frac{2\tilde{m}}{\Sigma_{II}(r)} - \frac{\tilde{\rho}_0 \tilde{\Sigma}_0^2 }{(1 - 2\omega)}\left(\frac{\tilde{\Sigma}_0}{\Sigma_{II}(r)}\right)^{2\omega}, & \text{if } \omega \neq \frac{1}{2}, \\
1 - \frac{2\tilde{m}}{\Sigma_{II}(r)} - \frac{\tilde{\rho}_0 \tilde{\Sigma}_0^3}{\Sigma_{II}(r)} \ln\left(\frac{\Sigma_{II}(r)}{\tilde{\Sigma}_0}\right), & \text{if } \omega = \frac{1}{2},
\end{cases}
\end{equation}
and,
\begin{equation}
B_{II}(r) = \frac{\Sigma_{II}'(r)^2}{A_{II}(r)}.    
\end{equation}
The prime in this expression stands for derivative with respect to the radial coordinate $r$. For concreteness, in our discussion we will focus on the case $\omega=1$, which corresponds to a Reissner-Nordström solution of charge $Q^2=\tilde{\rho}_0 \tilde{\Sigma}_0^2$ when $\Sigma_{II}\to  \tilde{\Sigma}_0 r$ but may develop a bounded internal region if $l_0>0$ or an unbounded one (wormhole) if $l_0<0$. 



\section{SCALAR PERTURBATIONS}\label{secIII}
Despite being a simplified model, scalar test field perturbations capture essential features of wave dynamics in curved backgrounds and are frequently used as a starting point for analyzing the stability and quasinormal modes of compact objects. Disregarding backreaction effects, let us consider a massless scalar perturbation governed by the Klein-Gordon equation
\begin{equation}\label{eq1}
\Box \Phi = \frac{1}{\sqrt{-g}} \partial_\mu \left( \sqrt{-g} g^{\mu\nu} \partial_\nu \Phi \right) = 0.
\end{equation}
Due to spherical symmetry, we assume the standard decomposition:
\begin{equation} \label{eq2}
\Phi( x^\mu) = \sum_{n,l=0}^\infty \sum_{m=-l}^l \frac{\Psi_{nl}(t, r) }{\Sigma(r)}Y_{\ell m}(\theta, \phi),
\end{equation}
where $Y_{lm}$ are the spherical harmonics. Substituting (\ref{eq2}) in (\ref{eq1}) one obtains that the $\Psi_{nl}(t,r^*)$ must satisfy
\begin{equation}\label{et}
\frac{\partial^2 \Psi_{nl}(t, r^*)}{\partial t^2} - \frac{\partial^2 \Psi_{nl}(t, r^*)}{\partial r^{*2}} + V_{eff}(r(r^*)) \Psi_{nl}(t, r^*) = 0,
\end{equation}
with,
\begin{equation}
V_{eff}(r(r^*)) = \frac{A'(r) \Sigma '(r)}{2 B(r) \Sigma (r)}-\frac{A(r) B'(r) \Sigma '(r)}{2 B(r)^2 \Sigma (r)}+\frac{A(r) \Sigma ''(r)}{B(r) \Sigma (r)}+\frac{A(r)l(l+1)}{\Sigma (r)^2},    
\end{equation}
where to eliminate the first derivative term, we considered the tortoise coordinate $r^{*}$, defined by
\begin{equation}
\frac{dr^*}{dr} = \sqrt{\frac{B(r)}{A(r)}}.    
\end{equation}
The literature presents several methods for solving this equation, with an excellent compilation available in \cite{Konoplya:2011qq, Pani:2013pma}. In the following, we introduce three independent approaches employed in this study.

\subsection{WKB METHOD}

In the study of quasinormal modes, the WKB method is used to obtain approximate analytical solutions of the quasinormal frequencies that solve Eq.(\ref{et}) subject to the boundary conditions $\lim_{r^*\to \pm \infty }\Psi_{nl}(t, r^*)\propto e^{-i \omega_{nl} t }e^{\pm i \omega_{nl} r^*}$, which represent damped waves emanating from the potential barrier (purely ingoing modes at the event horizon and purely outgoing at asymptotic infinity). 

The procedure consists in matching approximations to the wave equation which are valid in the asymptotic regions, near the event horizon ($r^*\to-\infty$) and at spatial infinity ($r^*\to+\infty$), to the local solution obtained by a Taylor expansion of the potential around its maximum, using connection formulas at the two turning points \cite{Konoplya:2011qq}. 
This method was first applied considering a third-order expansion \cite{Iyer:1986np}, later refined to sixth-order \cite{Konoplya:2003ii}, and is even available for thirteenth-order \cite{Matyjasek:2017psv}, though the sixth-order expansion is usually regarded as sufficiently good and is the one we will consider here. The QNM frequencies can thus be obtained 
as
\begin{equation}
\frac{i\left(\omega_{nl}^2 - V_0\right)}{\sqrt{-2V''_0}} - \sum_{i=2}^{6} \Lambda_i = n + \frac{1}{2},
\end{equation}
where $V_0$ is the potential at the maximum $r_0$ and $\Lambda_i$ are coefficients that depend on the effective potential and its derivatives at $r_0$. 

\subsection{POSCHL-TELLER METHOD}
In the method initially proposed in \cite{Ferrari:1984zz}, one approximates the effective potential by the function
\begin{equation}
V_{eff}(r(r^*)) \approx V_{\text{PT}}(r^*) = \frac{V_0}{\cosh^2(\beta(r^* - r^*_{0}))},
\end{equation}
where $V_0$ is the maximum of the effective potential and $\beta$ is given by
\begin{equation}
\beta^2 = -\frac{1}{2V_0} \left. \frac{d^2 V_{eff}(r(r^*))}{d{r^*}^2} \right|_{r^*=r^*_{0}}.
\end{equation}
With this new potential and considering a coordinate transformation, it is possible to obtain an analytical solution for equation (\ref{et}), expressed in terms of hypergeometric functions. The solution leads to
\begin{equation}
\omega_{nl} = - i \beta \left(n + \frac{1}{2} \right) \pm \sqrt{V_0 - \frac{\beta^2}{4}}.
\end{equation}
This approximation has yielded reasonable results in different geometries \cite{Berti:2009kk}.

\subsection{TIME DOMAIN}
Adopting the method developed in \cite{Gundlach:1993tp}, we define the null coordinates
$v \equiv t+r^*$ and $u \equiv t -r^*$, which allows us to rewrite Eq.(\ref{et}) in the form
\begin{equation}
\left( 4 \frac{\partial^2}{\partial u \partial v} + V_{eff}(u,v) \right) \Psi(u,v) = 0.
\end{equation}
Applying the finite difference method, one obtains
\begin{equation}
\Psi_N = \Psi_E + \Psi_W - \Psi_S - \frac{h^2}{8} V_{eff}(S)(\Psi_W + \Psi_E) + O(h^4),
\end{equation}
where $S = (u,v)$, $W = (u + h,v)$, $E = (u, v+h)$ and $N = (u+h,v+h)$, with $h$ the stepsize between two neighboring grid points. Given initial data on the null surface $ v = v_0$, which remain constant along $u=u_0$, the time evolution can be obtained iteratively. For concreteness, we will take a Gaussian wave packet 
\begin{equation}
\Psi (0,v) = A \text{exp}[-(v-v_c)^2/\sigma].   
\end{equation}
We will observe the field at $r^* = 10r_h$, with $h=0.1$, $A=1, v_c = 20$ and $\sigma = 9$. To find the dominant quasinormal modes that appear in the ringdown phase when the above wave packet interacts with the potential barrier, we can perform a fitting of the numerical integration data considering a linear expansion of the form
\begin{equation}
\Psi = \sum_{i=1}^n A_i e^{Im[\omega_i]t} \cos (Re[\omega_i]t + c_i),     
\end{equation}
which allows us to find the coefficients $A_i, \omega_i$ (real and imaginary parts) and $c_i$. 

\section{RESULTS}\label{secIV}

We now proceed to present the various configurations analyzed. We first begin by considering symmetric black bounces with horizons and asymptotically flat regions, and later we will address the asymmetric configurations.

\subsection{Symmetric black bounces}

\begin{figure}[!h]
\begin{center}
\begin{tabular}{ccc}
\includegraphics[height=5.0cm]{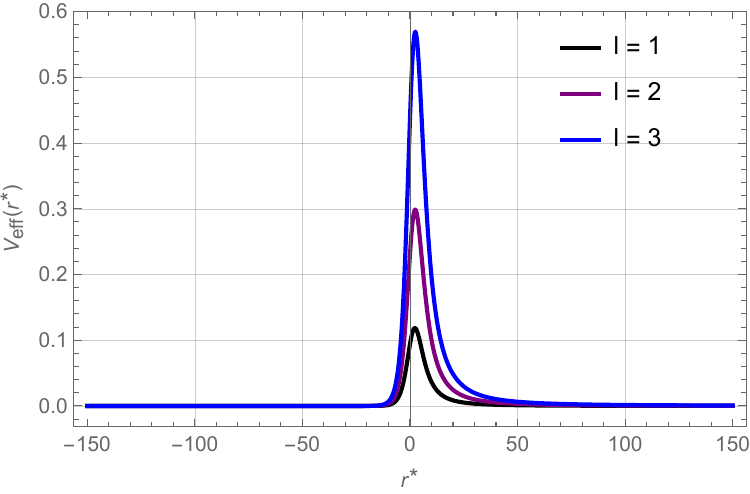} \includegraphics[height=5.0cm]{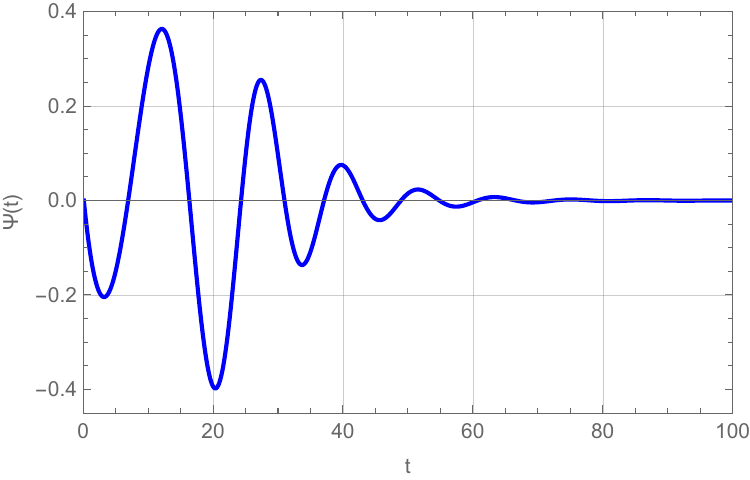}\\ 
(2a) \hspace{7.5 cm}(2b)
\end{tabular}
\end{center}
\caption{The effective potential (2a) and time evolution of  the massless scalar field (2b) for the symmetric black bounce of Eq. (\ref{symetric}) with $l=2, a = m = \omega = 1$ and $\rho_0= 0.5$. The observation point is located at $r^* = 100$. \label{Figure1}}
\end{figure}

On the left column of Fig. \ref{Figure1} we have (2a) the effective potential of a symmetric Kiselev solution for several values of angular momentum. The right column, plot (2b), represents the corresponding ringdown signal for the case $l=2$. This example represents a standard barrier potential with vanishing asymptotics both near the horizon and far from it. 


The corresponding quasinormal frequencies are listed in Table \ref{Table I}, from which we conclude that the fundamental mode oscillates more rapidly as $l$ increases. In contrast, the theoretical estimates for the decay rates (WKB and Poschl-Teller columns) decrease with increasing $l$. The time domain analysis, however, shows an anomalous decay in the damping of the $l=2$ mode as compared to $l=1$ and $l=3$ before decreasing for larger values of $l$. 

\begin{center}
\begin{table}[ht]
\centering
\caption{Fundamental quasinormal modes of the massless scalar field for the symmetric solution (\ref{symetric}) with $a= m = \omega =1$ and $\rho_0= 0.5$.}
\begin{tabular}{c||c||c||c}
\hline
\hline
$l$ &   WKB &  Poschl-Teller & Time-Domain \\
 \hline
 \hline 
1 & 0.321794 - 0.101983 i & 0.327420 - 0.103807 i & 0.307262 - 0.099533 i \\
2 & 0.531226 - 0.101819 i & 0.534570 - 0.102303 i & 0.522440 - 0.090411 i  \\
3 & 0.741366 - 0.101633 i & 0.743801 - 0.101826 i & 0.731209 - 0.099188 i \\
4 & 0.951175 - 0.092287 i & 0.955303 - 0.092356 i & 0.966082 - 0.093953 i \\
\hline
\hline
\end{tabular}
\label{Table I}
\end{table} 
\end{center}

\begin{figure}[!h]
\begin{center}
\begin{tabular}{ccc}
\includegraphics[height=4cm]{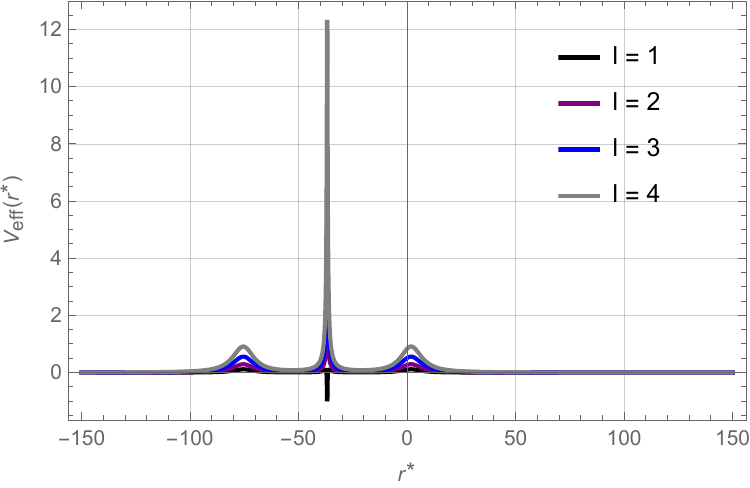} \includegraphics[height=4cm]{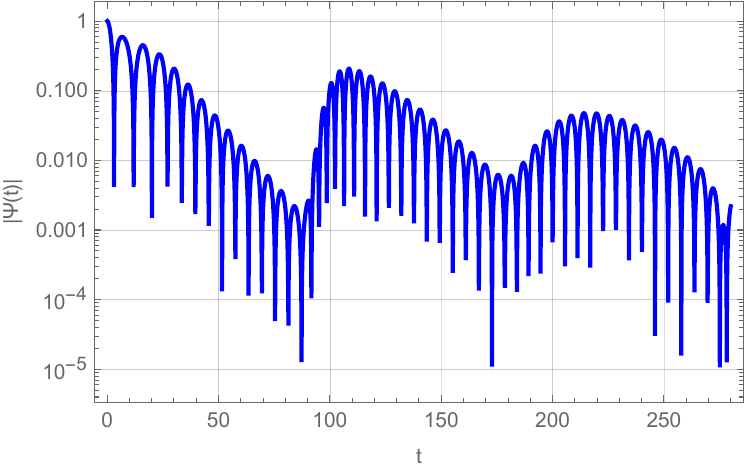}\\ 
(3a) \hspace{5 cm}(3b)\\
\includegraphics[height=4cm]{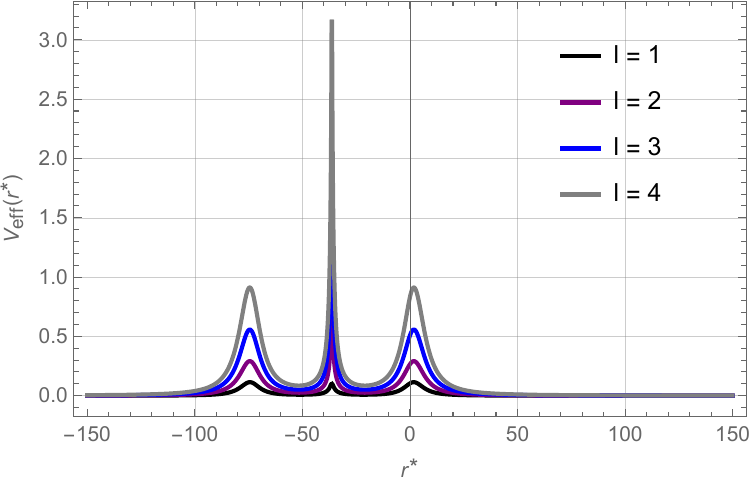} \includegraphics[height=4cm]{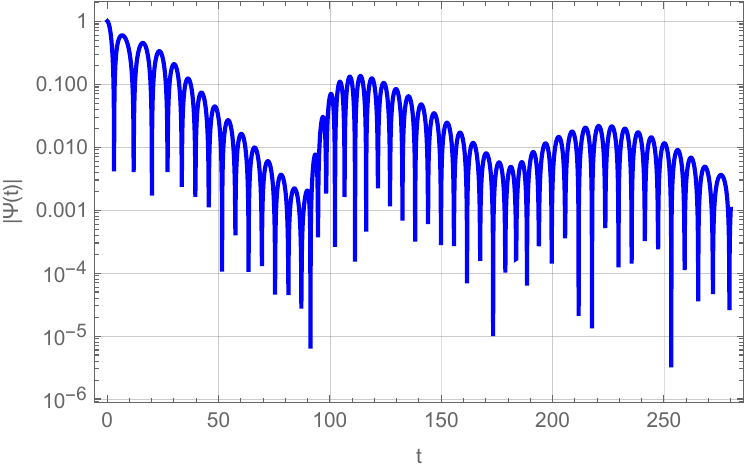}\\ 
(3c) \hspace{5 cm}(3d)\\
\includegraphics[height=4cm]{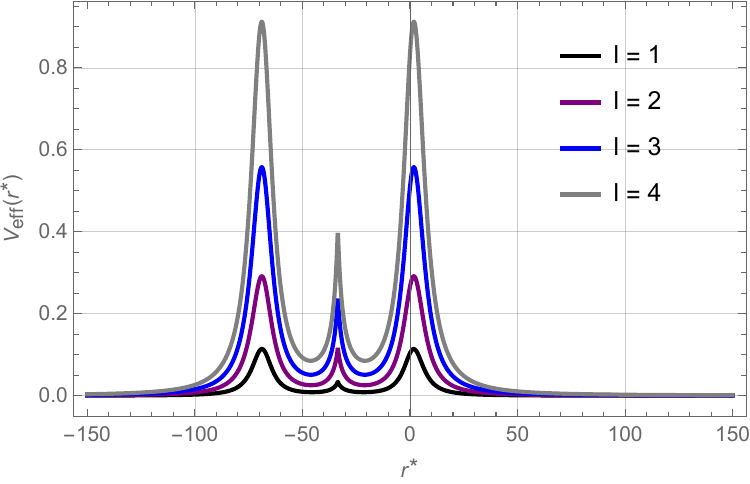} \includegraphics[height=4cm]{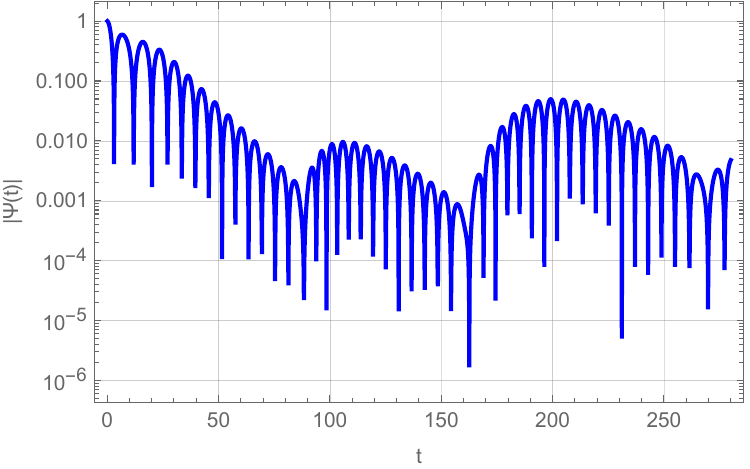}\\
(3e) \hspace{5 cm}(3f)\\
\includegraphics[height=4cm]{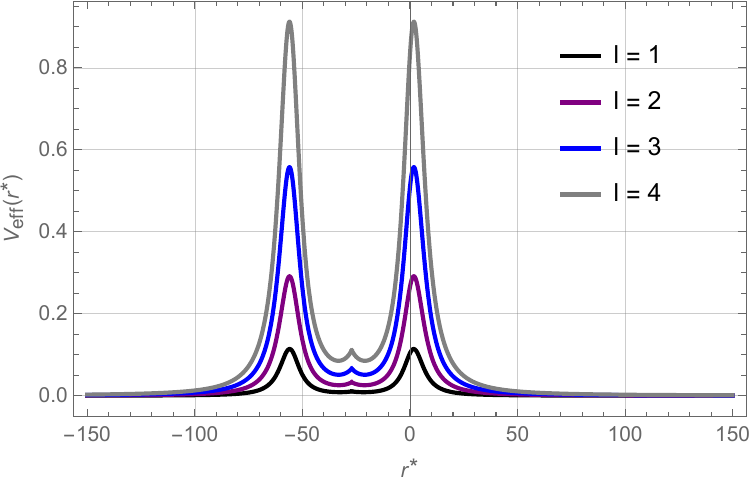} \includegraphics[height=4cm]{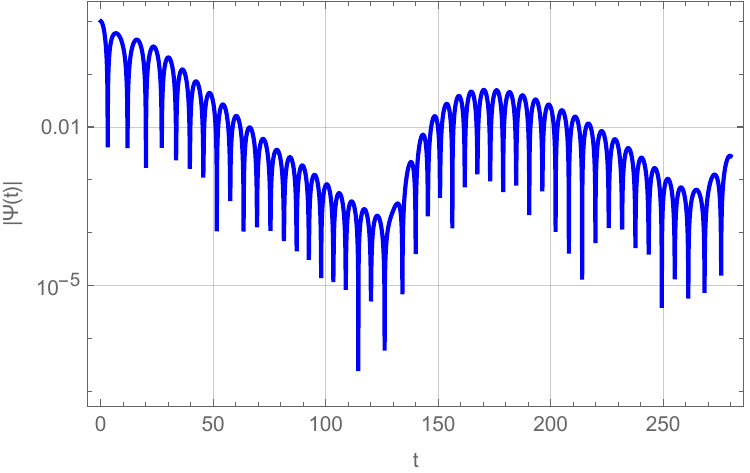}\\
(3g) \hspace{5 cm}(3h)
\end{tabular}
\end{center}
\caption{The effective potential (left panels) and time evolution of  the massless scalar field (right panels) for the symmetric solution (\ref{symetric}) with $\omega = 3/2, m = 1, \rho_0 = 65/27, l=2$, $ a = 0.8$ (3a), $a = 1$ (3c), $a = 1.2$ (3e) and $a  = 1.3$ (3g). The existence of echoes with a significant increase of the amplitude after several oscillations is evident.\label{Figure2}}
\end{figure}

Our second case of study corresponds to symmetric horizonless solutions with fixed constant paremeters but varying the minimal radius $a$. As can be seen in the left column of Fig. \ref{Figure2}, the effective potential is no longer a single barrier, showing more structure. In particular, besides having two centrifugal barriers on each side of the wormhole throat, for some model parameters a central barrier emerges at the throat whose amplitude can be much larger than the centrifugal parts [see (3a) and (3c)]. This third maximum of the potential is associated with an abrupt increase of gravitational redshift near the origin because $A(r) \sim \rho_0/a^3\gg 1$ when $a\to 0$. In these cases, the structure of echoes is clearly sensitive to the presence of the central peak, producing an increase in the amplitude of the oscillations at around $t\sim 80-90$. This effect progressively disappears as the central peak diminishes [see (3e) and (3g)] when the parameter $a$ increases. This clearly shows that the structure of echoes is strongly influenced by the number and height of peaks in the effective potential.  \\

\begin{figure}[!h]
\begin{center}
\begin{tabular}{ccc}
\includegraphics[height=5.0cm]{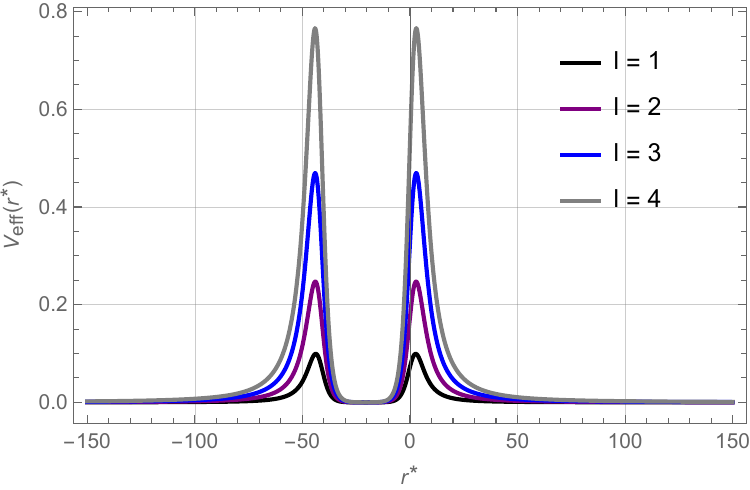} \includegraphics[height=5.0cm]{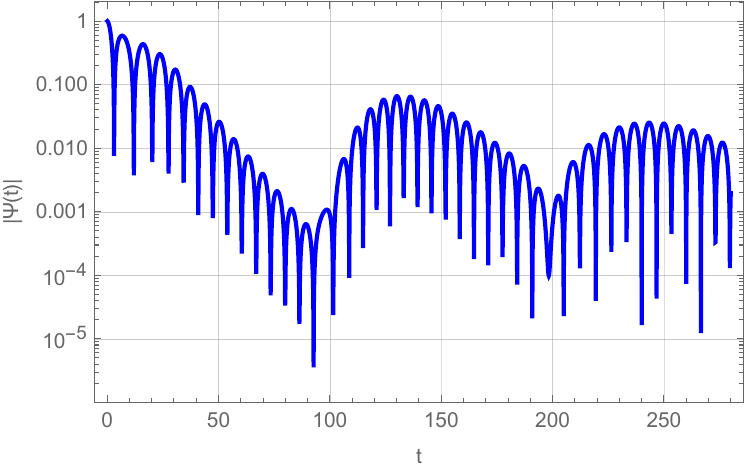}\\ 
(4a) \hspace{8 cm}(4b)\\
\includegraphics[height=5.0cm]{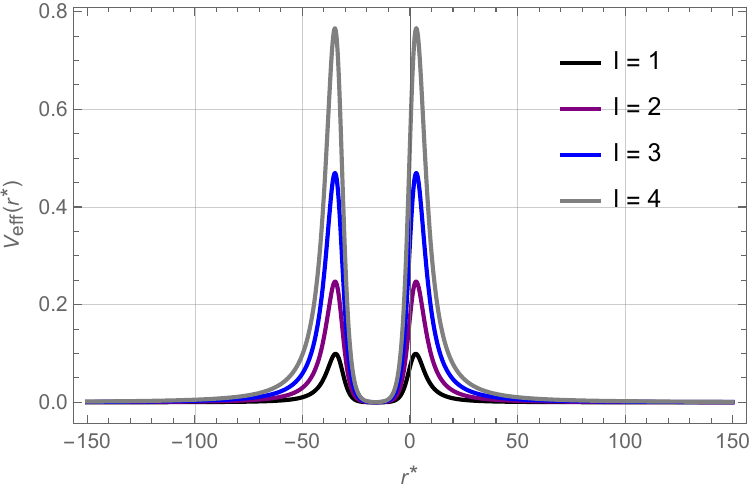} \includegraphics[height=5.0cm]{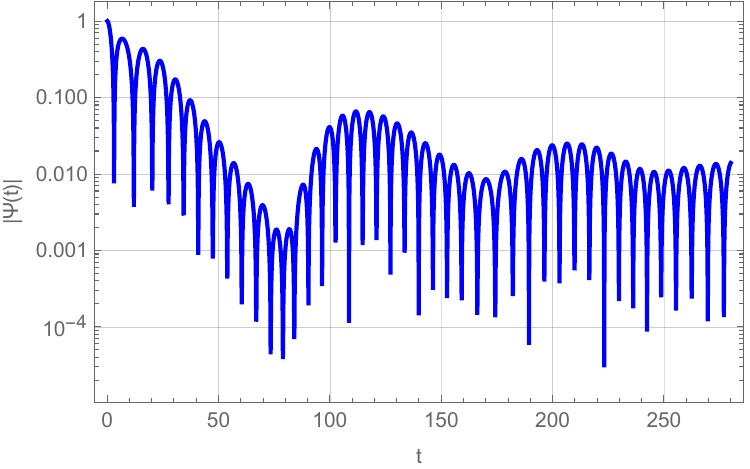}\\ 
(4c) \hspace{8 cm}(4d)\\
\includegraphics[height=5.0cm]{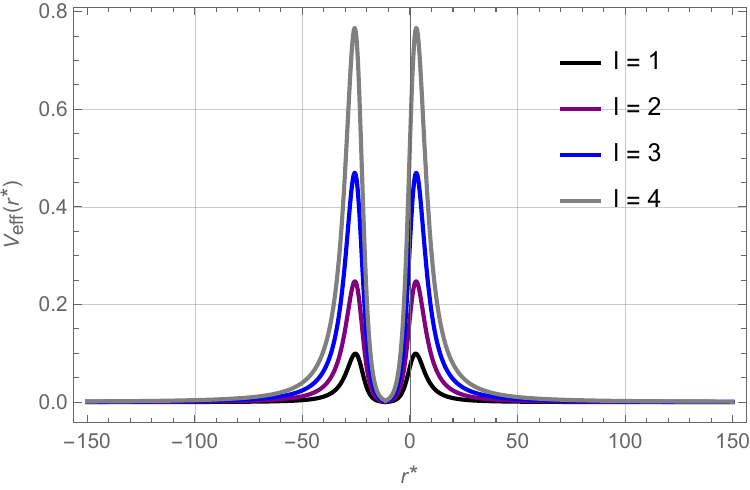} \includegraphics[height=5.0cm]{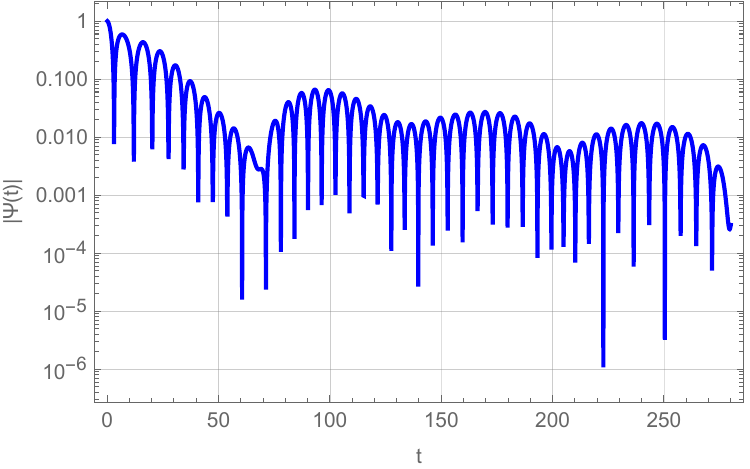}\\
(4e) \hspace{8 cm}(4f)\\
\end{tabular}
\end{center}
\caption{The effective potential (left panels) and time evolution of  the massless scalar field (right panels) for the symmetric solution (\ref{symetric}) with $l = 2, m = 1, \omega = 3/2, a = 2$, $\rho_0 = 10^{-4}$ (4a), $\rho_0 = 10^{-3}$ (4c) and $\rho_0 = 10^{-2}$ (4e). The existence of echoes is also evident in these cases.\label{Figure3}}
\end{figure}

In Fig.\ref{Figure3} we still consider the symmetric case but fix $a=2$ and vary the density parameter $\rho_0$. We find that the separation between the two barriers decreases as $\rho_0$ increases. This has a direct impact on the echoes because, as the barrier width diminishes, the temporal separation between successive echoes is reduced, yielding echoes that are more closely spaced in time and slightly attenuated.
 It is noteworthy that, as $\rho_0$ increases, the photon sphere shifts inward to smaller values of $r$, as can be verified from the equation of null geodesics
\begin{equation}
{\dot r}^2 = \frac{E^2}{B(r)A(r)} - \frac{L^2}{B(r)\Sigma^2(r)}, 
\end{equation}
and imposing the condition of a critical curve,  we obtain \cite{Guerrero:2022qkh}
\begin{equation}
\left. \frac{d}{dr}\left(\frac{A_I(r)}{ \Sigma_I^2(r)} \right) \right|_{r_{ps}} = 0,   
\end{equation}
which gives $r_{ps} (\rho_0 = 10^{-4}) = 2.23605, r_{ps} (\rho_0 = 10^{-3}) = 2.23588$ and $r_{ps} (\rho_0 = 10^{-2}) = 2.23420$, reflecting a greater concentration of mass–energy and the resulting enhancement of gravitational redshift. Note that, unlike black holes, the direct relation between shadows and quasinormal modes in horizonless configurations is not yet well established and requires further investigation \cite{Duran-Cabaces:2025sly,Koch:2025gaw}.\\

\subsection{Asymmetric black bounces}

\begin{figure}[!h]
\begin{center}
\begin{tabular}{ccc}
\includegraphics[height=5cm]{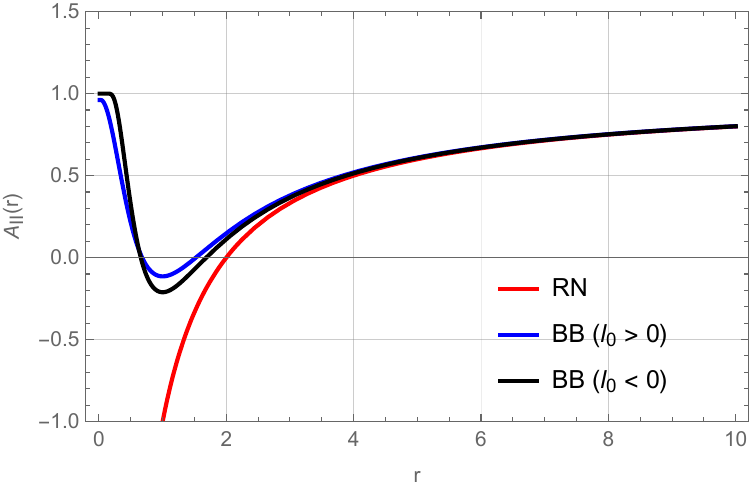} \includegraphics[height=5cm]{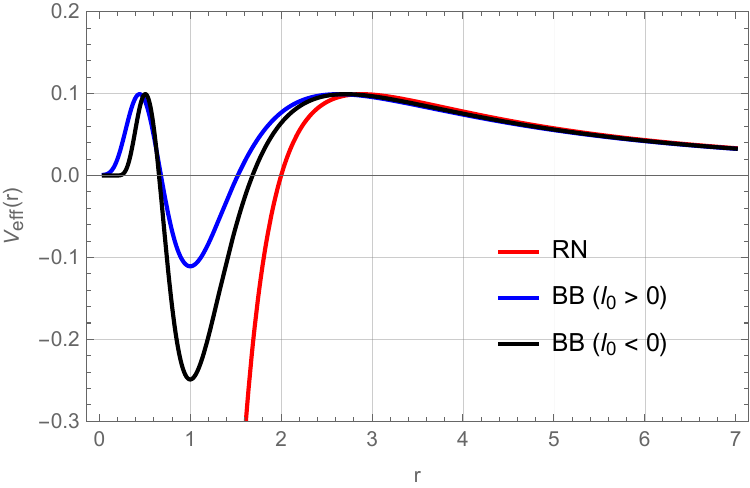}\\ 
(5a) \hspace{7 cm}(5b)\\
\includegraphics[height=5cm]{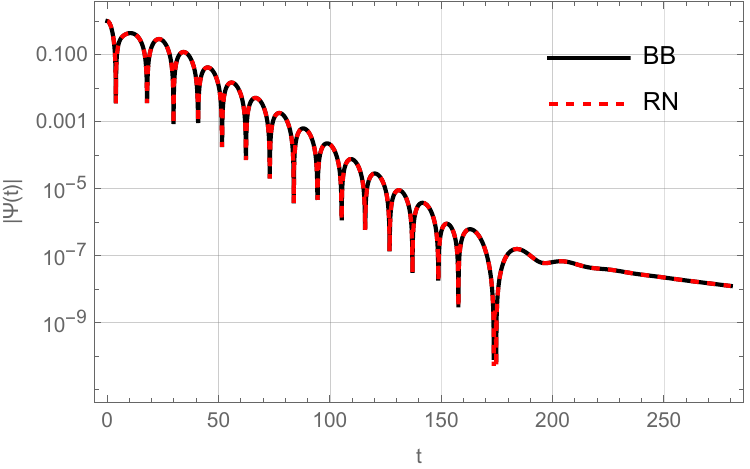}
\includegraphics[height=5cm]{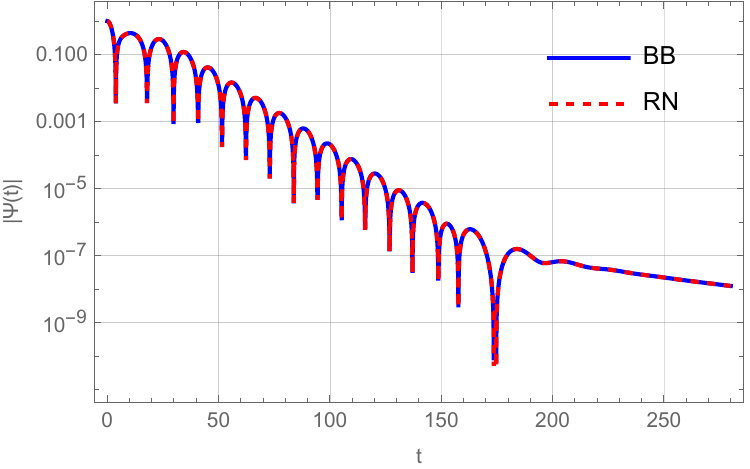} \\
(5c) \hspace{7 cm}(5d)\\
\end{tabular}
\end{center}
\caption{Graphical representation of $A_{II}(r)$ (5a), the effective potential for angular momentum $l=1$ (5b) and time evolution of the massless scalar field for the asymmetric solution (\ref{asymmetric}) for $l_0 = - 1/5$ (5c) and $l_0 = 1/5$ (5d) with  $\tilde{\rho}_0 = 0, \tilde{m} = r_0 = \omega = \alpha = 1$.\label{fig:asymBH}}
\end{figure}

\begin{figure}[!h]
\begin{center}
\begin{tabular}{ccc}
\includegraphics[height=5cm]{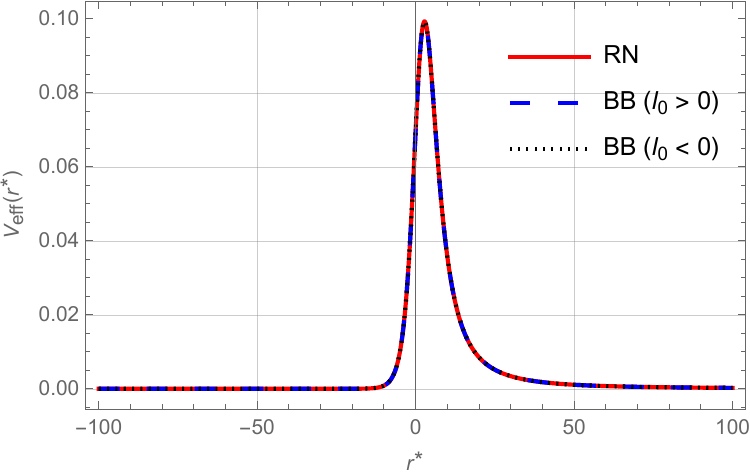} 
\includegraphics[height=5cm]{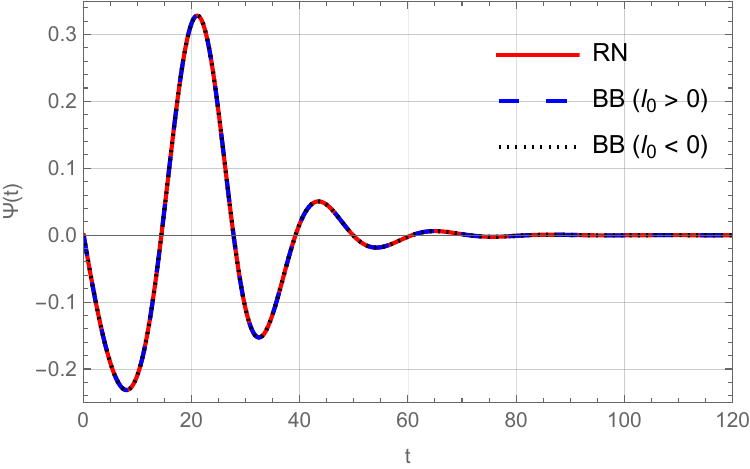}\\
(6a) \hspace{7 cm}(6b)\\
\end{tabular}
\end{center}
\caption{Effective potential (6a) and time evolution of  the massless scalar field (6b) for the asymmetric solution (\ref{asymmetric}) with $\tilde{\rho}_0 = 0, \tilde{m} = r_0 = \omega = \alpha = l = 1, l_0 = -1/5$ (black dotted line) and $l_0 = 1/5$ (blue dashed line).  The observation point is located at $r^* = 60$.\label{fig:6}}
\end{figure}

Let us now turn our attention to the  asymmetric solution (\ref{asymmetric}). We begin by considering black hole configurations (see Fig.\ref{fig:asymBH}). The first row of this figure, Figs.(5a) and (5b), represent the metric function $A_{II}(r)$ and its associated effective potential as a function of the radial variable $r$ for angular momentum $l=1$.   As one can see from Fig (5a), the red line represents a usual Schwarzschild black hole (no electric charge) with a single horizon at $r=2M$ (all lengths measured in units of $M$). Since the minimal $2-$sphere is located at $r_0=1$, the black bounce configurations exhibit two horizons. Though the location of the external horizon  is clearly different for the three configurations in (5a), the corresponding emission of scalar waves is essentially identical, as shown in Figs.(5c) and (5d). This is so because the relevant potential for the propagation of waves must be expressed in terms of a tortoise coordinate (see Fig.\ref{fig:6}), not in terms of $r$ as in (5b). The tortoise coordinate maps the region around the horizon into $r^*\to -\infty$, making the three resulting potentials become essentially identical. Thus, despite having completely different internal structures, the presence of an horizon prevents their identification via the emission of gravitational/scalar waves. The quasinormal frequencies that follow from the theoretical estimates and the time domain analysis appear in Table \ref{Table II}. As expected from Figs. (5c) and (5d), the frequencies do not depend on the sign of $l_0$. \\

\begin{center}
\begin{table}[ht]
\centering
\caption{Fundamental quasinormal modes of the massless scalar field for the asymmetric solution (\ref{asymmetric}) with $\tilde{\rho}_0 = 0, \tilde{m} = r_0 = \omega = \alpha = 1$, and $l_0 = \pm 1/5$. }
\begin{tabular}{c||c||c||c}
\hline
\hline
$l$ &   WKB &  Poschl-Teller & Time-Domain \\
 \hline
 \hline 
1 & 0.289717 - 0.0962071 i & 0.299026 - 0.098504 i  &  0.295918 - 0.093378 i \\
2 & 0.483256 - 0.0970068 i & 0.487328 - 0.098107 i & 0.477666 - 0.085006 i  \\
3 & 0.675233 - 0.0974105 i & 0.677867 - 0.097981 i & 0.666515 - 0.091291 i \\
4 & 0.867229 - 0.0975818 i & 0.869196 - 0.097926 i & 0.877047 - 0.092237 i \\
\hline
\hline
\end{tabular}
\label{Table II}
\end{table} 
\end{center}

\begin{figure}[!h]
\begin{center}
\begin{tabular}{ccc}
\includegraphics[height=5.0cm]{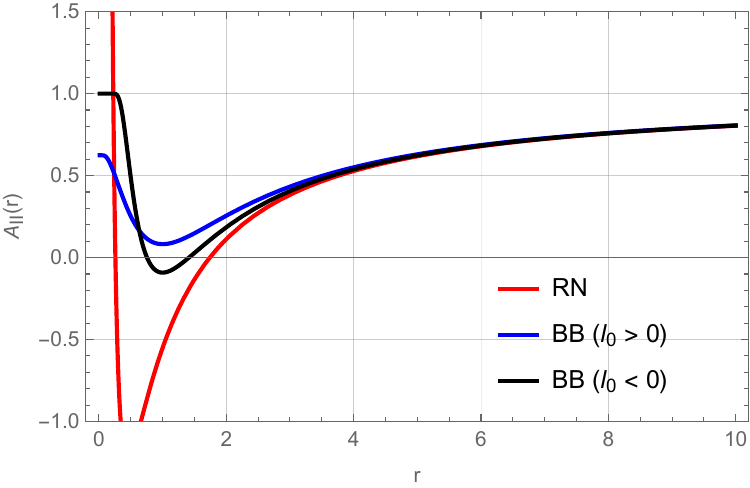} 
\includegraphics[height=5.0cm]{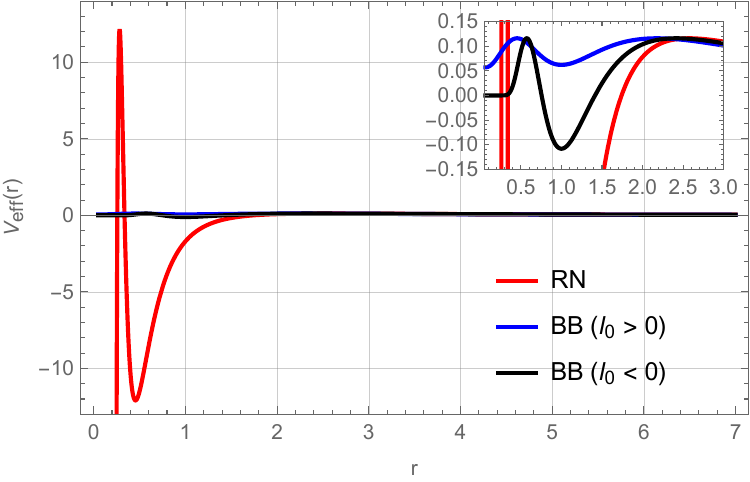} \\
(7a) \hspace{7 cm}(7b)\\
\includegraphics[height=5.0cm]{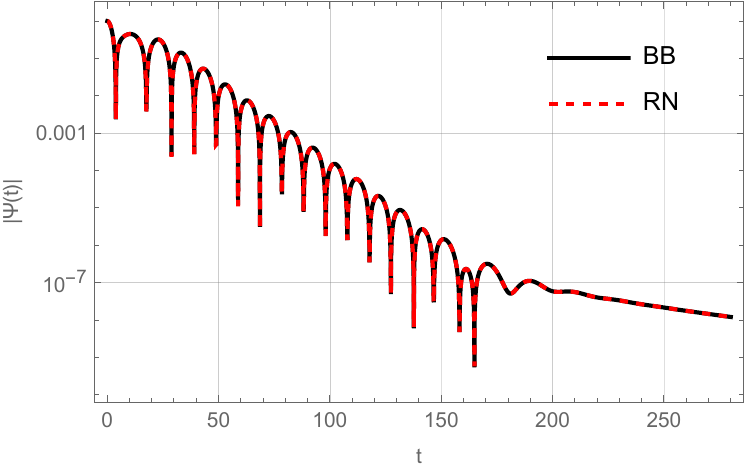}
\includegraphics[height=5.0cm]{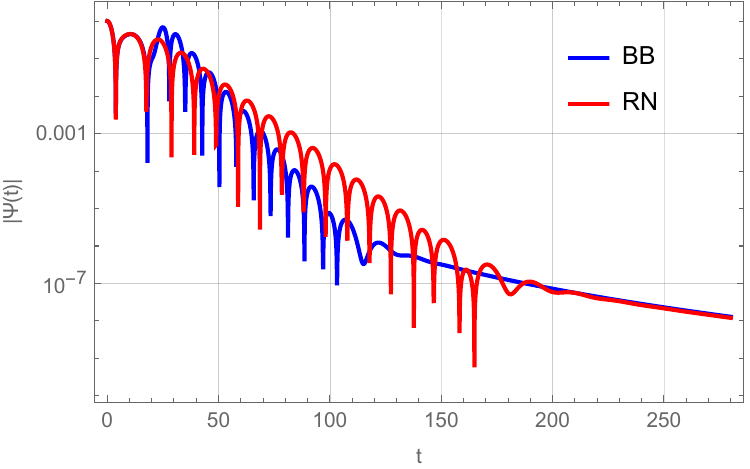} \\
(7c) \hspace{7 cm}(7d)\\
\end{tabular}
\end{center}
\caption{Graphical representation of $A_{II}(r)$ (7a), the effective potential (7b) and time evolution of the massless scalar field for the asymmetric solution (\ref{asymmetric}) for $l_0 = - 1/2$ (7c) and $l_0 = 1/2$ (7d) with  $\tilde{\rho}_0 = \frac{0.45}{\tilde{\Sigma}_0^{2 \omega +2}}, \tilde{m} = r_0 = \omega = \alpha = l = 1$.\label{Figure6}}
\end{figure}

\begin{figure}[!h]
\begin{center}
\begin{tabular}{ccc}
\includegraphics[height=5.0cm]{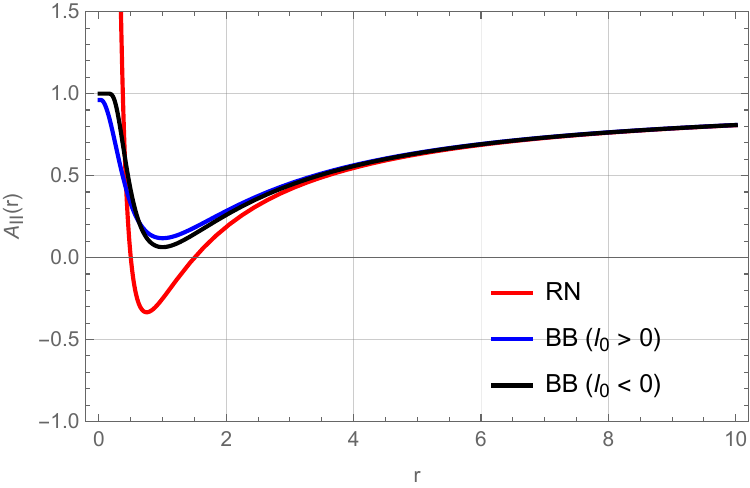} 
\includegraphics[height=5.0cm]{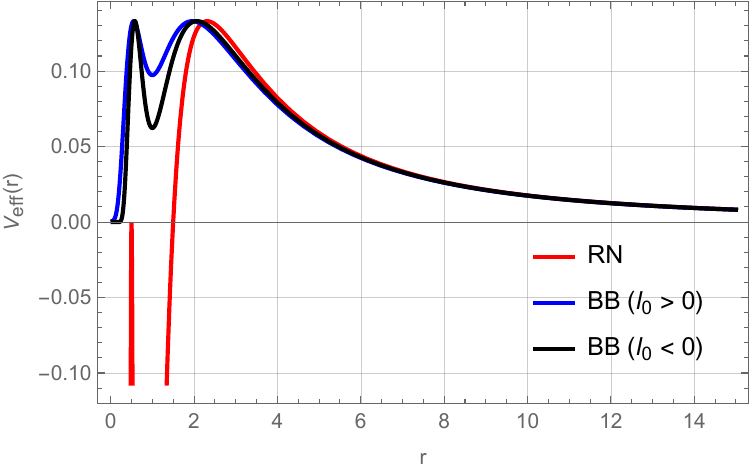} \\
(8a) \hspace{7 cm}(8b)\\
\includegraphics[height=5.0cm]{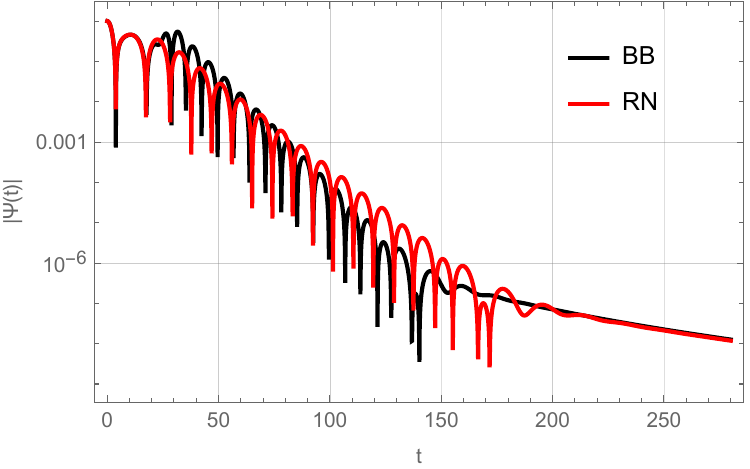}
\includegraphics[height=5.0cm]{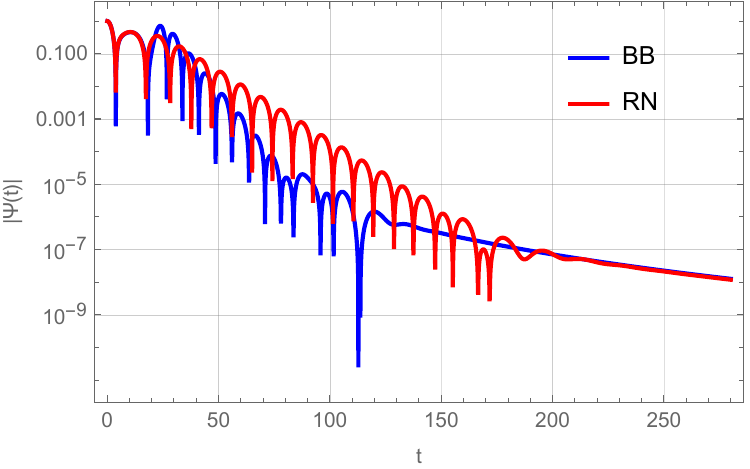} \\
(8c) \hspace{7 cm}(8d)\\
\end{tabular}
\end{center}
\caption{Graphical representation of $A_{II}(r)$ (8a), the effective potential (8b) and time evolution of the massless scalar field for the asymmetric solution (\ref{asymmetric}) for $l_0 = - 1/5$ (8c) and $l_0 = 1/5$ (8d) with  $\tilde{\rho}_0 = \frac{0.75}{\tilde{\Sigma}_0^{2 \omega +2}}, \tilde{m} = r_0 = \omega = \alpha = l = 1$.\label{Figure7}}
\end{figure}

\begin{figure}[!h]
\begin{center}
\begin{tabular}{ccc}
\includegraphics[height=4.0cm]{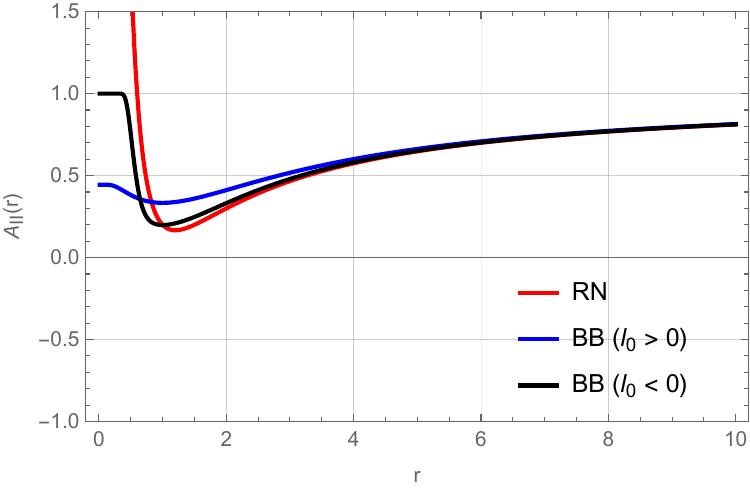} \includegraphics[height=4.0cm]{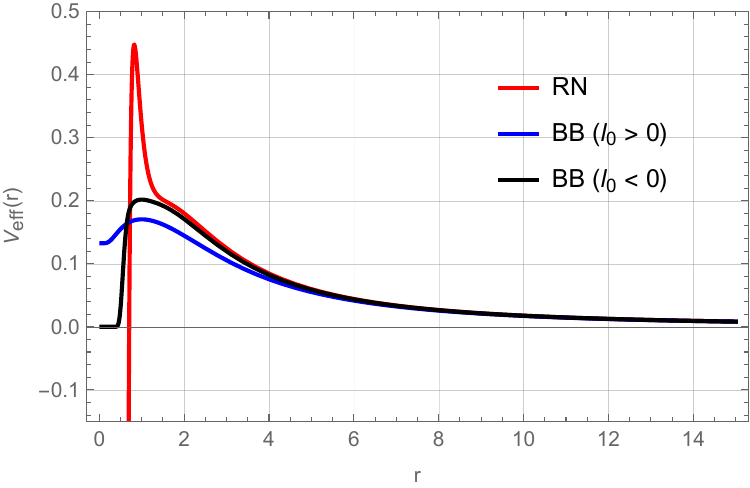}\\ 
(9a) \hspace{5 cm}(9b)\\
\includegraphics[height=4.0cm]{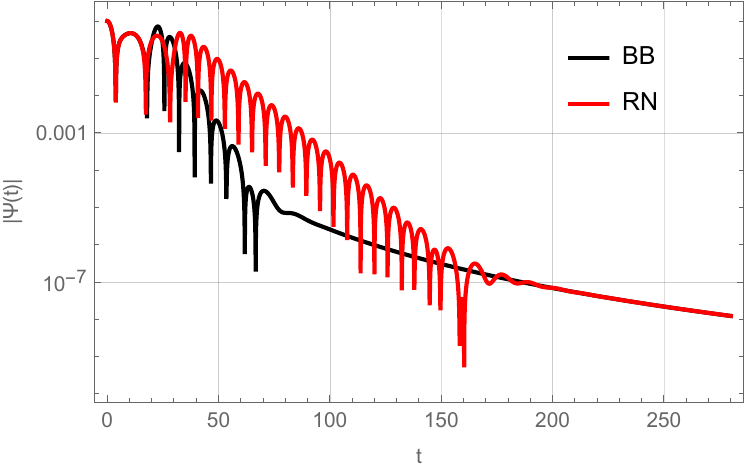} \includegraphics[height=4.0cm]{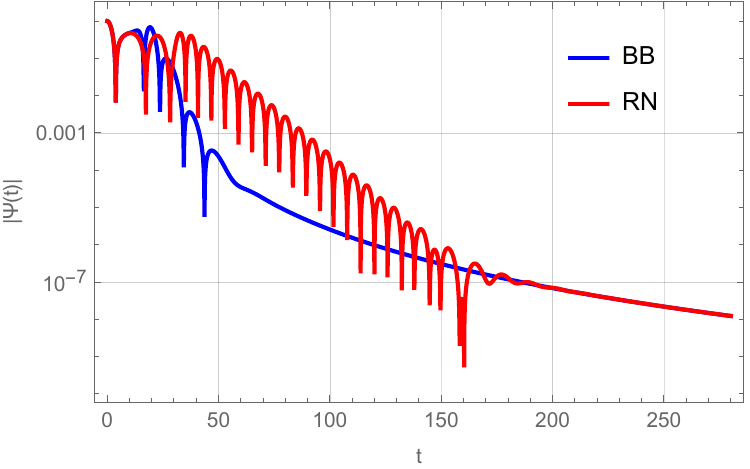}\\
(9c) \hspace{5 cm}(9d)\\
\includegraphics[height=4.0cm]{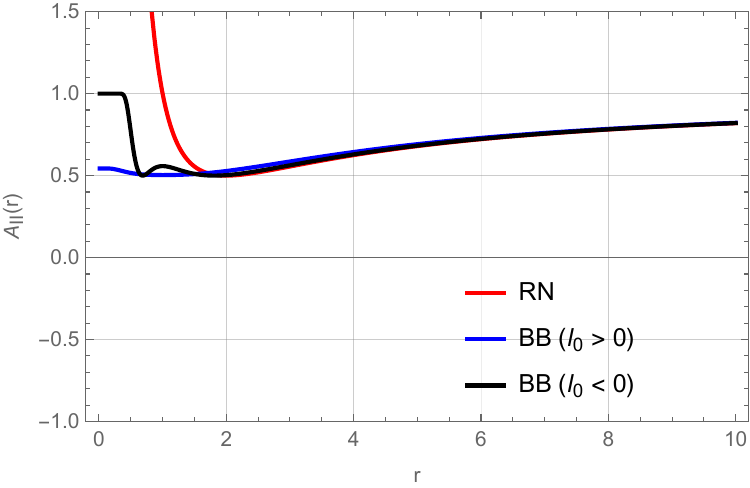} \includegraphics[height=4.0cm]{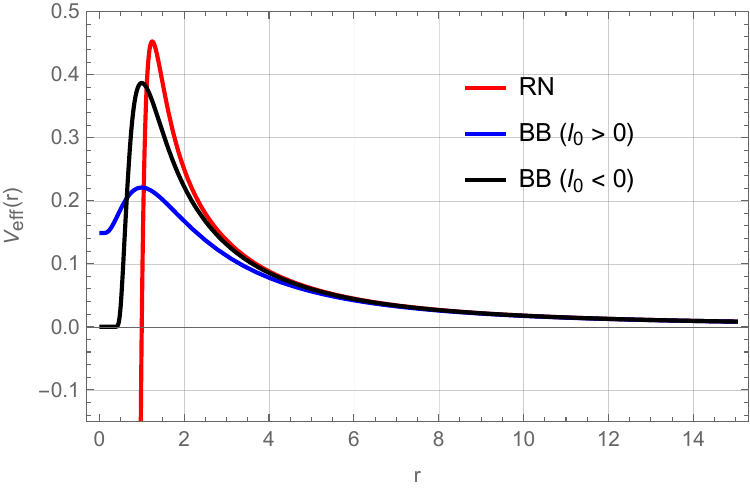}\\
(9e) \hspace{5 cm}(9f)\\
\includegraphics[height=4.0cm]{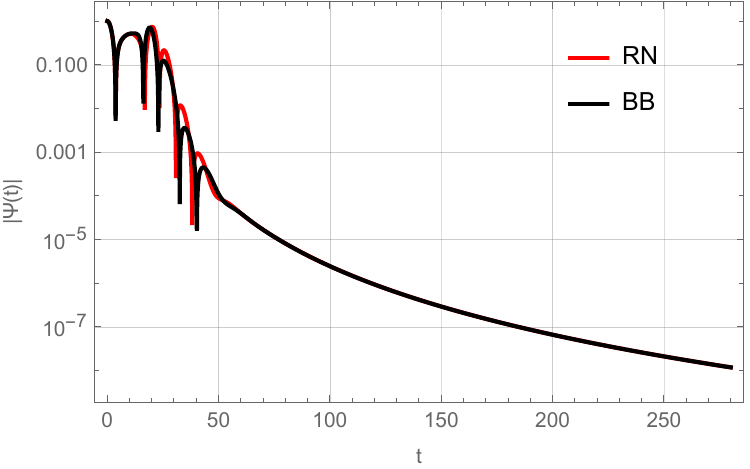} \includegraphics[height=4.0cm]{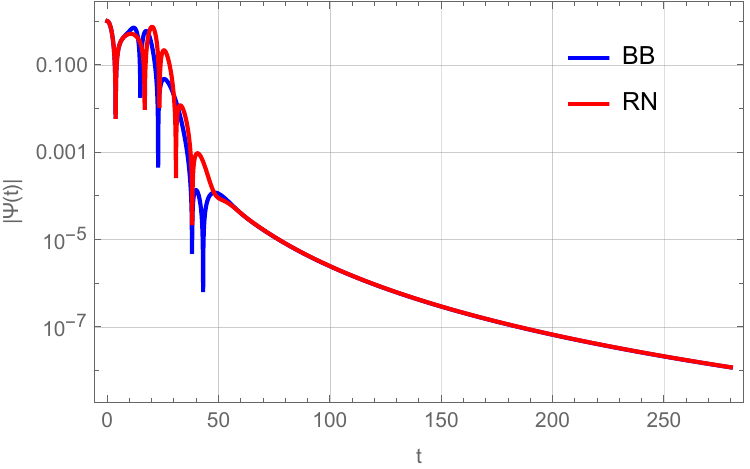}\\
(9g) \hspace{5 cm}(9h)
\end{tabular}
\end{center}
\caption{Graphical representation of $A_{II}(r)$ (9a), the effective potential (9b) and time evolution of the massless scalar field for the asymmetric solution (\ref{asymmetric}) with $\tilde{m} = r_0 = \omega = \alpha = l = 1, \tilde{\rho_0} = {1.2}/{\tilde{\Sigma}_0^{2 \omega +2}}, l_0 = - 1$ (9c) and $l_0=1$ (9d). Figures 9(e)–9(h) have the same parameters as in the first part but with $\tilde{\rho_0} = {2}/{\tilde{\Sigma}_0^{2 \omega +2}}$. \label{Figure9}}
\end{figure}

Let us now consider charged configurations, $\rho_0>0$, with and without horizons, as shown in Fig.\ref{Figure6}. For the chosen parameters, the unbounded black bounce ($l_0<0$) and the Reissner-Nordström configuration have horizons, while the bounded black bounce is horizonless. This has observable effects in the emitted waves, being the black hole emissions essentially identical, while the horizonless one exhibits different features. In particular, there is a slight increment in the amplitude of the waves after the first two oscillations, but then a faster decay and transit to the late-time power-law tail. This behavior seems to be the general trend for horizonless configurations, as shown in Fig.\ref{Figure7}. There we see that both black bounce cases are horizonless, their emission exactly fits the Reissner-Nordström profile during the first instants but then increases its amplitude to later decay faster and transition to the late-time tail much sooner than the Reissner-Nordström case. The difference between the bounded and unbounded cases is that the damping is more pronounced and the transition to the late-time power-law tail occurs earlier in the bounded case. For completely horizonless configurations, see Fig.\ref{Figure9}, the general trend observed in the previous cases is still present if the amount of charge is not too big. As charge is increased, the shape and height of the effective potentials (see Fig.(9f)) becomes more similar and the emission pattern rapidly degenerates, as shown in Figs. (9g) and (9h).

\section{Summary and conclusion}\label{secV}

We have studied the quasinormal mode spectrum of a massless scalar test field propagating on symmetric and asymmetric black bounce spacetimes with and without horizons. The geometries considered are sourced by anisotropic fluids in the framework of GR. The symmetric case coincides with a Kiselev spacetime where the radial variable experiences a shift of the type $r\to \sqrt{r^2+a^2}$, thus extending the domain of $r$ from $r\in [0,\infty[$ to $r\in ]-\infty,\infty[$. The asymmetric case represents two kinds of objects: a bounded universe (when $l_0>0$) that emerges when a minimal area surface is crossed, and an asymmetric wormhole representing an unbounded internal universe (when $l_0<0$). We have chosen the asymmetric case to be, essentially, a deformation of the Reissner-Nordström solution of GR, and manifests its new features around and below a certain $r$ located near the minimal area surface.  

Our results show that configurations with horizons exhibit a single-peaked effective potential and a standard ringdown waveform. The quasinormal frequencies obtained with the WKB and Poschl-Teller approximations in the symmetric configuration indicate that the oscillation frequency increases with the multipole number $l$, while the damping rate decreases, being in good agreement with the time-domain estimation (except for $l=2$). {In the asymmetric case, however, the WKB and Poschl-Teller methods yield a damping with opposite trends (growing in the former and decaying in the latter), while the time-domain results always yield smaller values, with an anomaly in $l=2$. Curiously, in the configurations with horizons the spectrum does not depend on the sign of $l_0$.} 

In contrast, horizonless symmetric solutions exhibit a potential with a richer structure,  characterized by the appearance of a third central barrier. This leads to the generation of intermediate gravitational wave echoes, whose spacing and amplitude are strongly influenced by the model parameters. In particular, we found that the third barrier becomes more pronounced as the parameter $a$ decreases, approaching the classical singular regime, and that the energy density $\rho_0$ controls the proximity between the potential peaks. As $\rho_0$ increases, the photon sphere shifts inward and the time interval between successive echoes becomes shorter, resulting in a more rapid echo sequence.

Another important outcome is the regularity of the effective potential in the asymmetric configurations in the $r\to 0^+$ limit (finite and non-zero when $l_0>0$ but vanishing when $l_0<0$). This clearly contrasts with the Reissner–Nordström effective potential, which diverges to negative infinity as the center is approached. The profile of the emitted waves when horizons are present are essentially degenerate and provide no information about the internal structure of the objects, which is natural given the restrictions imposed by the causal structure. Some effects begin to be observable when the horizon disappears, showing some generic differences between the Reissner–Nordström geometry and the regular black bounces, namely, a relative increase of the wave amplitude after the first few oscillations followed by a faster decay and earlier transition to the late-time power-law tail. This effect is more pronounced in the bounded case than in the wormhole case. Nonetheless, these effects occur at such low amplitudes that there is little hope for their observability even by next generation gravitational wave telescopes. If overextreme, horizonless configurations are compared with black bounces of identical asymptotic properties, see Fig.\ref{Figure9}, the general trend observed before is preserved for sufficiently low charge. For larger charges, all three configurations degenerate into the same emission profile, though the overlap is not as pronounced as in the black hole case. Interestingly, in all the asymmetric unbounded cases, which have a wormhole structure, we do not observe echoes in the emission profile, which indicates that not all wormholes have that characteristic feature.


{Our findings indicate that there might be scenarios in which the dominant quasinormal modes frequencies emitted by compact astrophysical objects may not be able to clearly convey relevant information about their inner structure, such as a singularity, a wormhole throat, or a bounded universe behind a throat. More in-depth analyses are thus necessary to quantify the range of frequencies necessary to fully identify such structures and the technical capabilities needed for their detection. Further work in this direction is currently ongoing as well as complementary tests that may help break the observed degeneracies.}

\begin{acknowledgments}
A.C.L Santos thank the Coordenação de Aperfeiçoamento de Pessoal de Nível Superior (CAPES), Grants no 88887.822058/2023-00 and 88881.983410/2024-01, for financial support and the Department of Theoretical Physics $\&$ IFIC of the University of Valencia- CSIC for the kind hospitality during the elaboration of this work. This work is supported by the Spanish National Grants PID2020-116567GB-C21 and PID2023-149560NB-C21, and the Severo Ochoa Excellence Grant CEX2023-001292-S, funded by MICIU/AEI/10.13039/501100011033 (“ERDF A way
of making Europe”, “PGC Generacion de Conocimiento”) and FEDER, UE. L.A Lessa would like to acknowledge Fundação de Amparo à Pesquisa e ao Desenvolvimento Científico e Tecnológico do Maranhão (FAPEMA), Grants FAPEMA BPD- 08975/24. R.V. Maluf would like to acknowledge Conselho Nacional de Desenvolvimento Científico e Tecnológico (CNPq), Grants PQ - 311393/2025-0.
\end{acknowledgments}

\end{document}